\documentclass[12pt,a4paper]{article}

\usepackage{amsmath}
\usepackage{amssymb}
\usepackage{mathrsfs}
\usepackage{graphicx}
\usepackage{dsfont}
\usepackage[pdftex,colorlinks=true,linkcolor=blue,citecolor=blue]{hyperref}
\usepackage{enumerate}

\numberwithin{equation}{section}

\begin{document}

\title{Eternal Black Holes and Superselection in AdS/CFT}

\author{
  Donald Marolf${{}^{1,2}}$\thanks{\href{mailto:marolf@physics.ucsb.edu}
    {marolf@physics.ucsb.edu}},~
  Aron C. Wall${{}^1}$\thanks{\href{mailto:aroncwall@gmail.com}
    {aroncwall@gmail.com}} \\ \\
  {\it ${}^1$Department of Physics, University of California at Santa Barbara,}\\
    {\it Santa Barbara, CA 93106, U.S.A.} \\
  {\it ${}^2$Department of Physics, University of Colorado,}\\
    {\it Boulder, CO 80309, U.S.A.} \\
  }

\date{\today}

\maketitle

\begin{abstract}
It has been argued i) that Lorentz-signature solutions with wormholes connecting $n$ asymptotically AdS regions describe bulk quantum states dual to $n$ entangled but non-interacting CFTs and ii) that such bulk wormhole states should be identified with similar entangled but non-interacting bulk systems, each describing quantum geometries with only a single asymptotic region.  But if the wormhole is to behave semiclassically, we show that conjecture (ii) cannot hold. Instead, the theory of asymptotically AdS bulk quantum gravity must admit superselection sectors with respect to the CFT observables that are labeled by the type of wormhole connections allowed between black holes.  Moreover, these superselection sectors are indistinguishable in the dual CFT.   Finally, we describe restrictions on the possible superselection sectors associated with the spin-statistics relation and the expectation that black holes lying in distinct asymptotically AdS regions may be approximated by well-separated black holes in a single asymptotically AdS region.
\end{abstract}

\newpage

\tableofcontents


\newpage

\section{Introduction}
\label{sec:intro}
It is well-known that general relativity admits black hole solutions with multiple asymptotic regions connected by wormholes.  Although one can debate whether such Lorentz-signature wormholes are `natural' or `physical,' at the classical level one is free to include them in the initial data \cite{Witt:1986ng}. The subsequent evolution of the spacetime is then determined.

The situation is much less clear at the quantum level.  Path integral formulations of quantum gravity suggest that one should sum over all manifolds with prescribed asymptotics, and thus that there should be some non-zero amplitude for wormholes to be created and destroyed \cite{Farhi:1989yr,Fischler:1989se,Fischler:1990pk}.  But since one expects the number of asymptotic regions to be fixed as a boundary condition, mono-asymptotic quantum gravity (with only a single asymptotic region) should be well-defined.  The mathematical tensor product of two mono-asymptotic theories is then a context with two asymptopia between which wormhole connections are forbidden by construction.  Other concerns involving creation of wormholes are similar to those arising in discussions of baby universe production or topology change more generally; see
e.g. \cite{Sorkin:1985bh,Witten:1989sx,Giddings:1988wv,Horowitz:1990qb,Fujiwara:1991hg,Carlip:1994tt,Barbon:2001di,Barbon:2002nw,Adams:2005rb,Ambjorn:2008we,Gibbons:2011dh}.

It is natural to ask if the
AdS${}_{d+1}$/CFT${}_d$ correspondence  \cite{Maldacena:1997re,Gubser:1998bc,Witten:1998qj} can shed light on this issue.  Early in the history of AdS/CFT it was noted
\cite{Maldacena:1998bw,Horowitz:1998xk,Balasubramanian:1998de} that CFT duals to solutions with $n$ asymptotically AdS regions\footnote{Times some compact manifold $X$ which will play no role in our discussion below.} would naturally involve $n$ entangled copies of the ``standard'' CFT (each of which will be denoted {\bf CFT}) defined on $S^{d-1} \times\mathbb{R}$. The restriction to any particular copy thus yields a mixed state \cite{Maldacena:1998bw,Horowitz:1998xk,Balasubramanian:1998de}. This idea was sharpened by Maldacena \cite{Maldacena:2001kr} who argued that the path integral naturally computes the CFT states dual to certain large eternal black holes.   In particular, \cite{Maldacena:2001kr} emphasized that while the $n$ copies of {\bf CFT} are entangled, each copy evolves with precisely the same dynamics as for $n=1$; there are no dynamical interactions of any sort between the distinct copies of {\bf CFT}.  As an example, for $n=2$ the maximal analytic extension of the AdS-Schwarzschild black hole is dual to a so-called thermofield-double entangled state of the form

\begin{equation}
\label{TFD}
|\psi \rangle = \sum_{n}  e^{-E_n/2T} |E_n\rangle_L |E_n \rangle_R,
\end{equation}
where $|E_n\rangle_{L,R}$ are energy eigenstates in two (left and right) copies of the standard CFT and $T$ is the Hawking temperature of the AdS-Schwarzschild black hole. We will say that the right-hand-side lives in {\bf CFT} $\times$ {\bf CFT}.

We refer to the idea that (at least some) Lorentz-signature solutions with wormholes connecting $n$ asymptotically AdS regions describe bulk quantum states dual to $n$ entangled but non-interacting CFTs as conjecture (i).  Further motivation for this conjecture comes from the relation between global AdS space and certain hyperbolic eternal black holes with non-compact horizons, as this relation leads immediately to a dual description of the form \eqref{TFD}.  A familiar example is AdS${}_3$ written in BTZ coordinates without identifications and keeping only the regions in the CFT spacetime that correspond to the usual boundaries of the BTZ black hole.   See also \cite{Skenderis:2009ju} for further discussion in AdS${}_3$.

Recent work has taken this idea somewhat further.  Since $|E_n\rangle_L$ should itself be dual to an energy eigenstate in some bulk theory with one asymptotic region,  the right hand side of \eqref{TFD} must be dual to a similar entangled state involving two dynamically independent copies of this mono-asymptotic bulk. As a result, Van Raamsdonk \cite{VanRaamsdonk:2009ar,VanRaamsdonk:2010pw} suggested that the wormhole state should be identified with an appropriate entangled sum of over product states involving two disconnected (quantum) bulk spacetimes\footnote{See \cite{Mathur:2010kx,Mathur:2011wg,Mathur:2012zp,Czech:2012be,Mathur:2012dx} for similar comments in related contexts.  There
are also connections to recent discussions of AdS/CFT in ``sub-regions'' of spacetime \cite{Czech:2012bh,Bousso:2012sj,Hubeny:2012wa,Bousso:2012mh}.}.  We refer to this idea as conjecture (ii).  One might then expect entanglement to somehow also play a key role in the dynamical formation (or destruction) of such wormholes.

We reconsider such issues below, focusing on physics inside the black hole.
We show in section \ref{eternal} that conjecture (ii) must fail if the wormhole interior is to behave semiclassically.  Instead, the theory of asymptotically AdS bulk quantum gravity should admit superselection sectors with respect to the CFT observables that are labeled by the type of wormhole connections allowed between black holes.  Moreover, these superselection sectors are indistinguishable in the dual CFT. Section \ref{multi} then describes
restrictions on the possible superselection sectors associated with the spin-statistics relation and the expectation that black holes lying in distinct asymptotically AdS regions may be approximated by well-separated black holes in a single asymptotically AdS region. We close with some further discussion in section \ref{disc}.  In particular, we address potential concerns regarding the assumed semiclassical physics in the wormhole interior.  An appendix investigates in more detail the manner in which an observer inside the black hole (say, one who entered from region $A$) can receive signals from another asymptotic regions $B$.

\section{The Eternal Black Hole}\label{eternal}

Recall the proposed identification \cite{VanRaamsdonk:2009ar,VanRaamsdonk:2010pw} of bulk wormhole states with two asymptotic regions and entangled sums of single-asymptotic region bulk states, termed conjecture (ii) above.
We argue below that this conjecture requires large violations of semiclassical physics inside the wormhole. More precisely, let $|{\rm w}2\rangle$ be any bulk state which approximates the spacetime with two asymptotic regions given by the maximal analytic extension of the (global) AdS-Schwarzschild black hole.   We will give a sense in which $|{\rm w}2\rangle$ is operationally distinct from any state $|{\rm TFD}\rangle$ in the direct product of two mono-asymptotic bulk theories.  Let us call the mono-asymptotic theory $\overline {\bf bulk}$ so that $|{\rm TFD}\rangle$ is a state in $\overline {\bf bulk} \times \overline {\bf bulk}$.
For definiteness we take $|{\rm TFD}\rangle$ to be the bulk dual to the `thermofield-double' state \eqref{TFD}.

We proceed by considering a gedankenexperiment involving some observer---let's call her Alice.  We will assume that 1) if Alice jumps across a black hole horizon and performs an experiment, the possible results of her experiment correspond to a single quantum operator in the usual way and 2) the semiclassical approximation is valid with high probability for classical objects moving around on large eternal black holes (far from the singularity).

Using the map between CFT sources and bulk boundary conditions\footnote{Some authors \cite{Kabat:2011rz,Bousso:2012sj,Czech:2012bh,Hubeny:2012wa,Kabat:2012hp} have suggested that local fields on one side of the bulk black hole are entirely built from operators in a single copy of {\bf CFT}.  We emphasize that our argument is not based on this conjecture.  In particular, we expect this conjecture to fail beyond leading order in the $1/N$ expansion due to difficulties in localizing observables in quantum gravity \cite{Giddings:2005id,Giddings:2007nu}; see in particular the discussion in section 4.2 of \cite{Heemskerk:2012mn}.}, we can with high probability create Alice near the (say, right) boundary by acting with a unitary operator $e^{iA}$ at Killing time $t_A$ in the right-hand copy ({\bf CFT}${}_R$) of {\bf CFT}.  We can also create another observer Bob near the left boundary using a second unitary operator $e^{iB}$ at Killing time $t_B$ in the left CFT ({\bf CFT}${}_L$). Note that $A$ and $B$ commute.    The resulting states are
$e^{i(A+B)} |{\rm w} 2\rangle$ and $e^{i(A+B)} |{\rm TFD}\rangle$.   We create both observers in such a way that they then plunge into the black hole.  Let $P$ be the operator projecting onto states in which Alice finds that she meets Bob inside the black hole.

\begin{figure}[h!]
  \centering
  \includegraphics[height= 8cm]{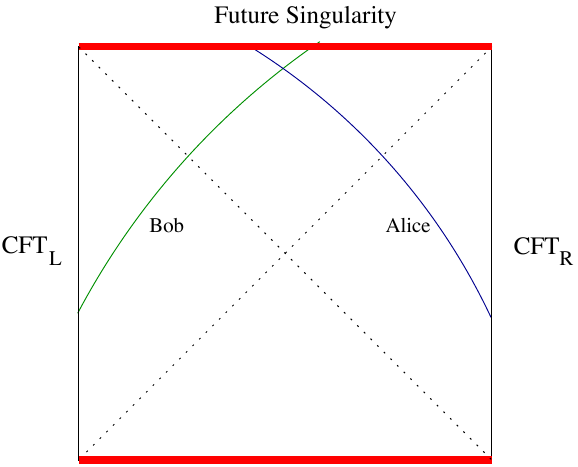}
  \caption{A conformal diagram showing Alice and Bob created on the right and left boundaries, falling into the black hole, and meeting inside. The dotted lines are horizons.}
    \label{ABfall}
\end{figure}

In a semiclassical wormhole, we can arrange for Alice and Bob to meet with high probability inside the black hole; see figure \ref{ABfall}.    So for appropriate $A$ and $B$ we have
\begin{equation}\label{w21}
\langle {\rm w} 2 | e^{-i(A+B)} P e^{i(A+B)} | {\rm w} 2 \rangle \approx 1 .
\end{equation}
But should we choose not to create Bob we have
\begin{equation}\label{w22}
\langle {\rm w} 2 | e^{-iA} P e^{iA} | {\rm w} 2 \rangle \approx 0.
\end{equation}

On the other hand, let us consider $e^{i(A+B)} |{\rm TFD}\rangle$.  Here Alice is created in
the right-hand bulk ($\overline {\rm \bf bulk}_R$).  By construction, this is a self-contained theory in its own right, and so contains an observable $P$ that describes whether or not Alice finds a Bob inside an appropriate black hole.  Since $\overline {\rm \bf bulk}_R$ is dual to {\bf CFT}${}_R$, this $P$ must depend only on {\bf CFT}${}_R$.  It follows that $P$ and $B$ commute so that
\begin{equation}
\label{11}
\langle {\rm TFD} | e^{-i(A+B)} P e^{i(A+B)} | {\rm TFD} \rangle = \langle {\rm TFD} | e^{-iA} P e^{iA} | {\rm TFD} \rangle.
\end{equation}
Identifying $|{\rm TFD}\rangle = |{\rm w} 2\rangle$ and using \eqref{w21} and \eqref{w22} would then imply the contradiction $1 \approx 0$.

We conclude that $|{\rm w}2 \rangle$ and $|{\rm TFD} \rangle$ are operationally distinct.  So if both are indeed dual to \eqref{11}, the two dualities must also be distinct in the sense that Alice's observations in the two bulk theories are described by different observables in {\bf CFT} $\times$ {\bf CFT}.  Said differently, composing the two dualities leads to a non-trivial duality between the wormhole theory (which we now denote $\overline {\rm \bf bulk}{}_{{\rm w}2}$) and $\overline {\rm \bf bulk} \times \overline {\bf bulk}$.

Some readers may find multiple bulk-to-boundary maps (and thus non-trivial bulk-to-bulk dualities) to be a surprising complication.  But this result is in fact natural from a variety of perspectives.  To frame the discussion, we introduce the notation {\bf bulk}${}_2$ (and similarly {\bf bulk}${}_n$) for the most general theory of bulk quantum spacetimes with two (or $n$) asymptotically AdS regions for which the algebra of boundary observables\footnote{One may think of this as the algebra generated by (rescaled) boundary values of bulk fields; see e.g. \cite{Marolf:2008mf,Marolf:2008tx} for a precise definition. Note that this algebra contains both local boundary fields and the Hamiltonian.   Since one expects general Wilson loops to mix with local operators under time-evolution in the dual CFT, the corresponding algebra in $\overline {\rm \bf bulk}$ should indeed be isomorphic to the full observable algebra in {\bf CFT}.} ${\cal A}_\mathrm{bndy}$ is isomorphic to the observable algebra of {\bf CFT} $\times$ {\bf CFT}.  In particular, we would like {\bf bulk}${}_2$ to contain as sub-theories both $\overline {\rm \bf bulk}{}_{{\rm w}2}$ and $\overline{\rm \bf bulk} \ \times \ \overline{\rm \bf bulk}$. More precisely, since the observables of each theory in {\bf bulk}${}_2$ contain ${\cal A}_\mathrm{bndy}$, the two theories above define different superselection sectors in {\bf bulk}${}_2$ with respect to ${\cal A}_\mathrm{bndy}$.

We may say that
{\bf bulk}${}_2 \ = $ {\bf CFT} $\times$ {\bf CFT} $\times$ $S_2$, where $S_2$ is a non-trivial space of superselection sectors.
In particular, the choice of state in $S_2$ determines whether or not distinct universes can be connected (and similarly for {\bf bulk}${}_n$ and $S_n$).
Ref. \cite{Marolf:2008tx} used state-counting to  argue for a result of precisely this kind in which the role of the second asymptotic region was played by a closed Friedman-Robertson-Walker (FRW) universe, and the role of our wormhole spacetime was played by Wheeler's `bag of gold' spacetime \cite{Wheeler}; see also \cite{Freivogel:2005qh} for related discussion.  Thus \cite{Marolf:2008tx} argued for superselection sectors associated with the possibility of wormholes in in {\rm bulk}${}_1$.

In fact, from a certain point of view the existence of superselection sectors in ${\rm \bf bulk}_1$ is manifest.
Consider a theory {\bf closed} describing a closed and connected FRW universe.  Now consider the formal product theory ${\rm \bf closed} \ \times \ \overline {\rm \bf bulk}$.  Under the definitions given above, this product theory lies in ${\rm \bf bulk}_1$  and every state of ${\rm \bf closed}$ defines a distinct superselection sector with respect to ${\cal A}_\mathrm{bndy}$.

Finally, recall that the description of the gravitational Hamiltonian as a boundary term can be used \cite{Marolf:2008mf} to argue that any bulk AdS quantum gravity theory is holographic in the sense that ${\cal A}_\mathrm{bndy}$ is a closed system that does not interact dynamically with any other degrees of freedom.  However, this argument does not rule out the existence of such additional degrees of freedom, which would then label superselection sectors\footnote{Ref. \cite{Marolf:2008mf} argued that there are no such superselection sectors at the perturbative level, but explicitly left open the possibility of non-perturbative superselection sectors of this kind.}.

Observables associated with $S_n$ may be said to parameterize bulk degrees of freedom which cannot be observed from the conformal boundary.  Since the Hamiltonian and other conserved momenta are themselves boundary observables, the superselection sectors $S_n$ must transform trivially under the AdS group. This means that, from the perspective of the conformal boundary, $S_n$ encodes information which is equally true at all times and places, similar to the alpha parameters of \cite{Coleman:1988cy,Giddings:1988wv}.  But an important distinction arises from our requirement that each sector in {\bf bulk}${}_n$ be dual to {\bf CFT}.  This implies that, apart from structure associated with superselection sectors in {\bf CFT} itself, each element of $S_n$ must be associated only with information not accessible from the boundary.  One expects that any such information is hidden behind a past or future horizon; otherwise an observer could in principle come in from the boundary, measure the information, and go back out to the boundary, which would affect the dynamics of the CFT.

It is interesting to explore the above thought experiment in more detail, and to understand
precisely how Alice might receive signals that she perceives as having come though the wormhole.  Certain facets of this issue are studied in appendix \ref{app}.  In particular we demonstrate that entanglement is not necessary for Alice to find a Bob that she perceives as coming through a wormhole.  Indeed, for $n=2$ there are $e^S$ disentangled states in which this occurs.  But entanglement turns out to be required if Alice is to encounter whatever kind of Bob we choose to throw into the black hole from the left boundary.

\section{A Puzzle with Multiple AdS regions}
\label{multi}

We now generalize these considerations to the theories {\bf bulk}${}_n$ associated with $n$ different asymptotically AdS regions.
\begin{equation}
\label{sectors}
{\rm \bf bulk}_n = ( {\rm {\bf CFT}} ) {}^n \times S_n.
\end{equation}
Because the $n$ CFT's do not interact, the operators in any two distinct CFT's must commute with each other (unless both operators are fermionic, in which case they anticommute).  Since the superselection sectors $S_n$ transform trivially under rotations of each CFT, by the spin-statistics relation the state of $S_n$ is always bosonic.

There is a permutation group $P_n$ exchanging the $n$ asymptotic regions, which therefore acts on ${\rm \bf bulk}_n$.  The action of $P_n$ on $( {\rm \bf CFT} ){}^n$ is manifest, although care must be taken with fermionic states because fermionic operators anticommute.  But how does $P_n$ act on $S_n$?   The possibility of nontrivial representations for $S_n$ leads to an interesting conclusion.  In particular, the following three reasonable sounding postulates are internally contradictory:

\begin{enumerate}
\item \label{worm} There exist states like $|{\rm w}2\rangle$ in {\bf bulk}${}_2$
whose physics has a well-defined semiclassical limit given by the maximal analytic extension of the Schwarzschild-AdS black hole, even for observers who fall into a black hole.

\item \label{product} For any two bulk theories (say with $n$ and $m$ asymptotic regions respectively), there is a direct product theory in which the two factors are strictly noninteracting; i.e., for any $n,m$ we have    {\bf bulk}${}_{n}$ $\times$ {\bf bulk}${}_{m}$  $\subset$ {\bf bulk}${}_{n+m}$.

\item \label{nto1} Excitations of a set of $n$ asymptotic regions can be approximated by $n$ widely separated excitations in a single asymptotic region.
\end{enumerate}

By Postulate \eqref{nto1}, we mean that for any state $|\psi\rangle_n$ in {\bf bulk}${}_n$, and for $g \in G_\mathrm{AdS}^n$ chosen so that the $n$ excitations are displaced by sufficiently large spatial distances,
there is a family of corresponding states $|\psi, g\rangle_1$ in {\bf bulk}${}_1$
whose physics is well-approximated by $|\psi\rangle_n$, except in the asymptotic regions as described below.
Here $G_\mathrm{AdS}$ is the AdS isometry group, and the precise requirements on $g$ will be described below.

The point of Postulate \eqref{nto1} is that, due to the asymptotically AdS boundary conditions, the physics near the AdS boundaries of any $|\psi\rangle_n$ in {\bf bulk}${}_n$ must approximate that of the product vacuum $\otimes_{i=1}^n |0 \rangle$ associated with the product spacetime $\times_{i=1}^n {\rm AdS}{}_d$. For a given $|\psi\rangle_n$, we can say that this approximation holds to some degree $\epsilon$ within some region ${\cal R}$ of the form
    \begin{equation}
    \label{outside}
    {\cal R} = \times_{i=1}^n {\rm AdS} /K,
     \end{equation}
     where $K$ is a set that we choose to be of the product form $K = \times_{i=1}^n K_n$ and whose intersection with any closed achronal surface is compact.  Using $g = (g_1,\dots,g_n) \in G_\mathrm{AdS}^n$ to identify each  AdS factor in \eqref{outside} with some canonical vacuum AdS spacetime, we can define the set $\tilde K = \cup_{i=1}^n g_i(K_i)$ and also $\tilde {\cal R} = {\rm AdS}/\tilde K$.  At least for $g$ large enough that the images of the $K_i$ do not intersect, we want there to be some $|\psi,g\rangle_1 \in {\rm \bf bulk}_1$ whose physics agrees with that of $|\psi\rangle_n$ (at least to an accuracy of order $\epsilon$ and for times much smaller than the AdS length scale) except for the replacement of ${\cal R}$ by $\tilde {\cal R}$.  Here we require $d > 3$, as any two BTZ black holes in AdS${}_3$ are always enclosed by a single common connected horizon.

The contradiction is reached as follows:   First consider $n=2$ asymptotic regions.  By Postulate \eqref{worm}, there exist semiclassical wormhole states in some superselection sector of {\bf bulk}${}_2$.  Any such superselection sector must correspond to some state of $S_2$.  Under the action of $P_2$, this state is either  a) symmetric, b) antisymmetric, or c) a superposition of states of the form (a) or (b), in which case we can choose to project onto either case.

Now choose $n = 3$ AdS regions, labeled $A, B, C$.
By Postulate \eqref{product}, there must exist superselection sectors of {\bf bulk}${}_3$ in which $A$ and $B$ can be connected by a semiclassical wormhole throat while $C$ is completely disconnected from either $A$ or $B$.  By symmetry, there also exist superselection sectors in which $B$ and $C$ are the connected regions, or $A$ and $C$.  All three sectors are operationally distinct by reasoning similar to the $n = 2$ case of section \ref{eternal}.
The 3 possibilities correspond to a three dimensional subspace of $S_3$ which breaks up into irreducible representations (irreps) under the action of $P_3$.  This symmetry group has three possible irreps: $\mathbf{1_S}$, the totally symmetric (trivial) irrep; $\mathbf{1_A}$, the totally antisymmetric irrep, and $\mathbf{2}$, the 2 dimensional irrep.   When the $S_2$ state is described by case (a) we have $\mathbf{2 + 1_S}$, while case (b) yields $\mathbf{2 + 1_A}$.

Once we know that there exist superselection sectors in nontrivial representations of $P_n$, we can choose the superselection sector and the CFT states independently.  Let us choose the state of $({\rm \bf CFT} ){}^3$ to be in the pure disentangled state $|\psi \rangle |\psi \rangle |\psi \rangle$ (i.e. each CFT is in the same state $|\psi \rangle$).  By applying Postulate \eqref{nto1}, one ends up with a state with three identical objects at large spatial separation in the same AdS region.  But identical objects with integer spin must have bosonic statistics, and objects with half-integer spin have fermionic statistics.  In neither case can they transform in $\mathbf{2}$.  Similarly, for superselection sectors in $\mathbf{1_A}$, one would have particles with the wrong kind of (bosonic or fermionic) statistics for their spin.

Note that the statistics of identical objects can be measured by an outside observer using interference experiments, and hence communicated to the boundary. It follows that nontrivial $P_3$ irreps would lead to differing physical dynamics for the boundary CFT observables.  This contradicts the assumption that the AdS/CFT duality is exact in each superselection sector.

Which of the three postulates should be rejected?  It would be surprising if Postulate \eqref{worm} were wrong, because it would require a seemingly valid semiclassical state (the eternal black hole) to be inconsistent for deep quantum gravity reasons.  However, those who believe in large violations of locality, even in semiclassical situations, might choose to reject it.  One would then resolve the puzzle by saying that when Alice jumps into the black hole, what she sees can depend \emph{only} on the state of her own CFT.  This might be a consistent way to eliminate superselection sectors from AdS/CFT.

At one level, Postulate \eqref{product} is just a mathematical fact.  One has to be able to take tensor products of independent systems at the mathematical level.  No one can stop us from considering the tensor product of a copy of QCD in 4 dimensions with a copy of $\phi^4$ theory in 2 dimensions, although this combined system is of no physical relevance.  Thus, if we assume the eternal black hole is consistent, then it is simply an indisputable mathematical fact that different superselection sectors of  {\bf bulk}${}_n$ exist, involving many possible representations of $P_n$.  This looks bad for Postulate \eqref{nto1}.

However, there may be a subspace of superselection sectors in which Postulate \eqref{nto1} holds but which contains no direct products of the sort described by Postulate \eqref{product}.  All such superselection sectors would be completely symmetric under $P_n$.  Because of this neutrality under permutations, in each superselection sector either a) no wormhole connections between asymptotic AdS regions are allowed or b) wormhole connections between any two asymptotically AdS regions are allowed.

Of course, the fact that a given sector allows connections between any two AdS regions need not imply that every state actually exhibits all such connections -- and certainly not with a macroscopically large wormhole.  Consider in particular the case $n=2$ and a state \eqref{TFD} in which a wormhole connects two black holes of entropy $S$. By acting with a time translation in, say, {\bf CFT}${}_R$ but not {\bf CFT}${}_L$ we can generate further states describing otherwise-identical black holes connected in different ways (i.e., with a relative time-translation) through an otherwise-identical wormhole. Assuming that the energy levels have incommensurate frequencies, this generates a space of $e^S$ states given by all superpositions of the form $|E \rangle |E \rangle.$  But we see no reason why there should be significantly more states describing similar wormholes in similar black holes.  In particular, the number of states associated with perturbative excitations is well-known to be much smaller than $e^S$.
Since there are $e^{2S}$ states containing a black hole of entropy $S$ in each region, the fraction of black hole states describing large wormholes may be as small as $e^{-S}$.

In contrast, given two large black holes, each in its own asymptotic region, the fraction of states in which they are connected by a small wormhole should be much larger.  This is readily argued by starting with large black holes of energy $E_0$, temperature $T$, and entropy $S_0$ in each region and taking them to be in generic states (thus likely not containing wormholes).  If there is also a tiny two-asymptotic region wormhole (of the rough form $|{\rm w}2\rangle$ dual to \eqref{TFD}) of energy $E_w$,  and if this wormhole then merges with both of the original black holes, the semiclassical result is a space of at least $e^{S_0}$ states in which a small wormhole now connects the two large black holes.  For small enough wormholes, the relation $dE = TdS$ implies that this is a fraction $e^{-E_w/T}$ of the full set of black hole states with energy $E_w + E_0$.

\section{Discussion}
\label{disc}

Our work above argued for superselection sectors in the AdS/CFT correspondence associated with the existence of wormhole connections between black holes.  Some sectors would allow such wormhole connections, while others would forbid them.  The issue arises for any number $n$ of asymptotically AdS regions and in particular for $n=1$, where the basic picture agrees with that described in \cite{Marolf:2008tx}.  As a refinement of this picture, we also discussed restrictions on the $n=1$ superselection sectors associated with the spin-statistics relation and the corresponding likely failure (at least in some superselection sectors of {\bf bulk}${}_n$) of the idea that black hole states with $n$ asymptotic regions should be similar (at least for short times) to states involving widely separated black holes in a single asymptotically AdS region.  In particular, we argued that sectors allowing wormhole connections between any particular two black holes must in fact allow connections between arbitrary pairs of black holes.

This conclusion leads to interesting issues if one allows an exponentially large number of black holes of order $C e^{S}$, where $S$ is the Bekenstein-Hawking entropy of a typical black hole under discussion.  Let us suppose that the superselection sector allows wormhole connections.
Even if the probability of two particular black holes being connected by large wormholes is of order $e^{-S}$, the large number of black holes implies a large probability for each black hole to be connected to {\it some} other black hole.  For $C \gg 1$ the typical number of connections will be large, and an observer falling into any black hole will likely see large and complicated signals arriving from such wormholes as soon as they cross the horizon.  The idea that such signals are large suggests that they lead to the observer's destruction.  So when these enormously large numbers of black holes are present, and in sectors where wormholes are allowed, it would appear that observers entering black holes do not experience smooth horizons.  One is tempted to ask if similar issues arise for cosmological horizons in the late stages of inflation (where general relativity predicts exponentially many Hubble volumes), though we refrain from pursuing this question in detail here.\footnote{One may also ask about Rindler horizons in flat space, though the infinite entropy of Rindler horizons provides a greater degree of protection.}

Our analysis has focused on very coarse properties of wormholes, such as whether they can exist at all.  It is interesting to ask if other properties of wormhole geometries may also be superselected.  For example, suppose we start with the standard wormhole $|{\rm w} 2\rangle$ with a throat of some radius $r_0$.  By coupling our CFTs to appropriate auxiliary systems, we could arrange to add energy to both sides of the wormhole so that ${\bf CFT} \times {\bf CFT}$ ends up in any state we desire.  In particular, we could choose it to be the state $|{\rm w} 2\rangle$ associated with a throat of radius $2r_0$.  But semi-classically this addition of energy does not increase the size of the original throat; indeed, the throat is hidden behind the past horizons of the original black hole and cannot be affected by any material that enters from infinity.  This raises the possibility that there is some throat-size observable (e.g. the geometry of an extremal co-dimension 2 surface inside the throat) whose values are also superselected.

We have focused above on the consequences of an observer (Alice, who falls into the black hole from region $A$) receiving signals from another asymptotic region ($B$).    Some initial steps toward understanding the role of entanglement in this process were taken in appendix \ref{app}.  Interestingly, we found many \emph{dis}entangled states ($e^S$ for $n=2$) in which Alice meets a Bob that she perceives as coming through a wormhole.   However, entanglement \emph{is} necessary to have a wormhole state in which Alice will encounter whatever Bob we choose to create on the left boundary (say, with the same overall center of mass motion directed accurately into the wormhole, but with an arbitrary internal quantum state).

The arguments in this work require the wormhole spacetimes to behave semiclassically, even inside the horizon.  While this assumption is natural due to the weak spacetime curvatures, and while it is unclear what conjectures (i) and (ii) might mean if it were to fail, we cannot exclude its violation by novel quantum gravity effects. Indeed, it was recently argued in \cite{Almheiri:2012rt} that some novel effect should invalidate our use of the semiclassical approximation near the horizon for generic quantum states of black holes with a single asymptotic region\footnote{
Some violation of this approximation is guaranteed by \cite{Mathur:2009hf}, though the question of how this would affect our semiclassical observers was left open.  The fuzzball proposal (see e.g. \cite{Lunin:2001jy,Mathur:2005zp,Mathur:2008nj}) might also invalidate our analysis.
While the fuzzball complementarity suggested in \cite{Mathur:2010kx,Mathur:2011wg,Mathur:2012dx,Mathur:2012zp} predicts that semi-classical physics would correctly describes the experiences of any Alice and Bob heavier than $T_H$, the predictions when Bob is an $E \sim T_H$ photon are less clear.}.  The status of this argument is less clear when the black hole connects to multiple asymptotic regions\footnote{With $n > 1$ asymptotic regions a given infalling observer can probe only one connected component of the black hole horizon (connecting the black hole to one of the asymptotic regions).  This opens the possibility that outgoing Hawking modes in (say) the left asymptotic region might be identified with what an observer entering the black hole from the right would describe as the internal partners to the outgoing Hawking modes in the right asymptotic region (see \cite{Susskind:2012uw} for an explicit statement, though the idea seems to be implicit in many discussions of black hole complementarity \cite{Susskind:1993if,Susskind:1993mu}). Similarly, in the case of a black hole which forms from collapse in a mono-asymptotic region, the firewall could be averted if the information in the early Hawking radiation is identified with the internal partners of the late Hawking radiation.  However, \cite{Susskind:2012uw} argues that this version of complementarity is inconsistent, because nothing prevents the early Hawking radiation from being sent back into the black hole, thus meeting itself and violating the no-cloning theorem in a single causal region; see also \cite{RB}.  A similar paradox can be arranged for the two-asymptotic-region black hole, though it requires  temporarily coupling {\bf CFT}${}_L$ at early times to {\bf CFT}${}_R$ at late times so that information radiated out of the right black hole is sent into the one on the left.}.  But we again emphasize that the large wormholes studied here appear to be exponentially rare.  So even if generic multi-asymptotic-region black holes contain firewalls, states describing large enough wormholes could well remain smooth.

{\it Note added in proof:}
The argument in section 2 assumed that $\overline {\bf bulk}$ is a self-contained theory in the sense that it can answer all questions about its bulk observers. In particular, it was assumed to contain a projection $P$ that describes whether or not Alice finds a Bob inside an appropriate black hole.  We also assumed
$\overline {\bf bulk}$ to be dual to  {\bf CFT}.  It is an interesting question whether both assumptions can in fact hold simultaneously; i.e., whether a sufficiently large set of quantum measurements inside black holes can in fact be described by boundary observables.  The larger theory {\bf bulk}${}_1$ is by definition self-contained and so certainly contains such a $P$.  But if $S_1$ in \eqref{sectors} is non-trivial then the projection $Q$ from
{\bf bulk}${}_1$ to $\overline {\bf bulk}$ is also non-trivial and could fail to commute with $P$.  Nevertheless, since trivial $S_1$ would imply $Q = {\mathds 1} $, the argument suffices to show that $S_1$ is non-trivial as desired.

We also mention that while the standard use of the term superselection sector would require the algebra of observables in $S_n$ to be abelian, we explicitly allow them to be non-abelian here.

\medskip

\noindent{\bf Acknowledgements:} It is a pleasure to thank
Joe Polchinski, Mark van Raamsdonk, and William Donnelly for interesting discussions involving black holes with wormholes. This work was supported in
part by the National Science Foundation under Grant Nos PHY11-25915
and PHY08-55415, and by funds from the University of California. Additional support came from FQXi grant RFP3-1008 for DM  and from the Simons Foundation for AW.  DM also thanks the University of Colorado, Boulder, for its hospitality during the final stages of this work.

\appendix

\section{Quantum Communication through a Wormhole}

\label{app}

This appendix will explore how quantum signals can be communicated in the context of an eternal black hole, from {\bf CFT}${}_L$, to an observer (Alice) who jumps into a black hole from {\bf CFT}${}_R$.  In particular, we will show that for any particular choice of Bob, there exist at least $e^S$ disentangled states in which Alice sees Bob.  However, if we want to be sure that Alice can meet \emph{any} Bob created with the same overall center of mass motion, it is necessary for the state to be entangled (as it is in \eqref{TFD}).  The fact that Alice and Bob can communicate quantum information implies the existence of a ``quantum teleportation'' protocol \cite{Bennett:1992tv}, which requires entanglement.

\paragraph{No entanglement needed to meet a particular Bob.}
Consider the superselection sector associated with the eternal black hole.  Each CFT is described by a mixed state with entropy $S$, which is entangled with its thermofield-double so as to make a pure state $|{\rm w}2\rangle$.  This mixed state corresponds to an ensemble of pure states peaked around the classical value of the energy $E_h$ and angular momentum $J_h$.  (We will use the term ``angular momentum'' broadly to mean all other operators in the global symmetry algebra that commute with the Hamiltonian, including R-symmetries as well as rotations.)

If Alice and Bob jump into the black hole with properly related overall center of mass motions, then Alice should meet Bob with high probability $p \approx 1$.  What are the necessary conditions for Alice to find Bob inside the black hole?  We will show that there is in fact a very large number of states (more than $e^S$) in which Alice observes a Bob.  Furthermore, in each of these states the two CFT's are unentangled.

Let $P$ be the projection operator corresponding to ``Alice meets some particular Bob'' (which is logically distinct from the question of whether Bob meets some Alice, since there may be states with additional copies of Bob or Alice besides those created on the boundary).  Assuming, as is natural, that $P$ is defined relationally with respect to the right-hand boundary only, $P$ must be invariant under the AdS group of the {\bf CFT}${}_L$.

Let $A$ and $B$ be unitary operators that create an Alice or a Bob at time $t = 0$ at the first or second boundary respectively.  Let Bob have an energy and angular momentum peaked around the classical values $(E_B,\,J_B)$, and let $S_f$ be the final entropy of the black hole, as viewed from {\bf CFT}${}_L$, after dropping in Bob and waiting for everything to thermalize.  We will make two assumptions about the dynamics of the CFT:
\begin{enumerate}
\item Nondegeneracy: no two states have exactly the same quantum numbers of energy $E$ and angular momentum $J$.
\item Ergodicity: When Bob jumps into the black hole, every state with energy $E_h + E_B$ close enough to $E$ and angular momentum close enough to $J_h + J_b$ has a nonzero probability to appear.  ``Close enough'' means within the range of uncertainty of $E$ and $J$ respectively.  The final entropy $S_f$ is simply given by the logarithm of the number of states within this range (up to subleading logarithmic terms which we will neglect).  We will assume that this Ergodicity property holds even after projecting with $P$ onto the states in which Alice meets Bob.
\end{enumerate}
Then it follows that there are at least $e^{S_f}$ distinct states $|\chi_n \phi_n \rangle$ which are factorizable (no entanglement between the two CFT's) such that $P$ is true (Alice is certain to meet Bob):
\begin{equation}
\langle \chi_n \phi_n | P | \chi_n \phi_n \rangle = 1.
\end{equation}

To show this, we use $|\psi \rangle$ as in \eqref{TFD} and $A,B$ as in section \ref{eternal} and write the state $Pe^{i(A+B)}| \psi \rangle$ using an energy eigenbasis on {\bf CFT}${}_L$:
\begin{equation}\label{Esum}
Pe^{i(A+B)}|\psi \rangle = \sum_{n = 1}^{e^{S_f}} c_n | \chi_n E_n \rangle
\end{equation}
where the $c_n$'s are complex coefficients, and the $E_n$'s are energy eigenvalues lying within the range of uncertainty of $E$.  The $\chi_n$'s are an overcomplete basis of $e^{iA}\mathcal{V}$, where $\mathcal{V}$ is the $e^S$-dimensional vector space of black hole states in {\bf CFT}${}_R$, whose energy is within the range of uncertainty of $E_h$.  The Ergodicity property guarantees that each of the $c_n$'s is nonzero.

Since the choice of superselection sectors carries no energy or angular momentum, $P$ commutes with $E$ and $J$.
 This means it also commutes with any function $f(E)$, such as a delta function projecting onto a single energy eigenvalue of $E$, and a complete commuting set of quantum numbers for $J$.  Using the Nondegeneracy property, it follows that Alice meets Bob in each of the terms of Eq. (\ref{Esum}) taken separately:
\begin{equation}
P|\chi_n E_n \rangle = |\chi_n E_n \rangle,
\end{equation}
each of which is a disentangled state.  (In the case of energy eigenvalues related by a global symmetry, the $\chi_n$'s for each state are the same.)  Consequently, there exist at least $e^{S_f}$ orthogonal factorizable states in which Alice meets Bob.  This shows that the ability of Alice to meet this particular Bob does not depend on the existence of a quantum entangled state.

\paragraph{Entanglement needed to meet any Bob.}
In order to regard a state as a wormhole, it is necessary that Alice be able to measure, not just one particular Bob created on the boundary, but \emph{any} Bob created on the boundary, at least if Bob jumps into the black hole in the right way to meet Alice, and Bob is light enough not to significantly perturb the wormhole metric.

In general, Bob contains quantum information, which is specified by the quantum operator which creates Bob in {\bf CFT}${}_L$.  For example, Bob can carry a qubit into the black hole, where the initial state of the qubit depends on the operator which creates Bob in {\bf CFT}${}_L$.  However, since Alice's measurements are invariant under the left-hand AdS group, it follows that Alice can only measure the AdS irrep of Bob's CFT (besides observables in her own CFT).  But by the Nondegeneracy Property, the set of observables in a CFT that commute with $E$ and $J$ is a commuting subalgebra.  This creates a conundrum: AdS/CFT says that Bob can only communicate \emph{classical} bits of information to Alice, but the semiclassical wormhole approximation says that Bob can communicate qubits to Alice.

The resolution of this conundrum is that the semiclassical wormhole approximation is only valid in certain states, in which there is a large amount ($S$) of entanglement entropy between the two CFT's.  It is an interesting fact of quantum information theory that in order to communicate a qubit, it is sufficient if the sender and receiver have a pair of entangled qubits, and can communicate two classical bits of information.  This can take place using the ``quantum teleportation'' protocol of \cite{Bennett:1992tv}.  In the absence of entanglement, classical communication cannot be used to reconstruct a qubit; consequently the consistency of the AdS/CFT eternal black hole must implicitly rely on quantum teleportation!


\addcontentsline{toc}{section}{Bibliography}
\bibliographystyle{JHEP}
\bibliography{./selectbibliography8}

\end{document}